\documentclass[a4paper,fleqn,usenatbib]{mnras}
\usepackage[T1]{fontenc}
\usepackage{ae,aecompl}
\usepackage{epsfig}
\usepackage{graphicx}
\usepackage{epstopdf}	
\usepackage{amsmath}	
\usepackage{amssymb}
\usepackage{footnote}
\usepackage{longtable}
\usepackage{supertabular,booktabs}	
\usepackage{ulem}
\usepackage{gensymb}

\title{Radiative properties of Cu-isoelectronic As, Se and Br ions for astrophysical applications}

\author[Jyoti et al.]{
Jyoti,$^{a}$
{Harpreet Kaur$^a$,}
{Bindiya Arora$^{a}$}\thanks{bindiya.phy@gndu.ac.in}
and
{B. K. Sahoo$^{b}$}\thanks{bijaya@prl.res.in}
\\
$^{a}$Department of Physics, Guru Nanak Dev University, Amritsar, Punjab-143005, India\\
$^{b}$Atomic, Molecular and Optical Physics Division, Physical Research Laboratory, Navrangpura, Ahmedabad-380009, India}

\begin{document}

\maketitle

\begin{abstract}
We present precise radiative data of line strengths, transition probabilities and oscillator strengths for the allowed transitions among the 
$nS_{1/2}$, $nP_{1/2,3/2}$, $n'D_{3/2,5/2}$ and $n'F_{5/2,7/2}$ states with $n=4$ to $6$ and $n'=4,5$ of the Cu-isoelectronic As, Se 
and Br ions. Due to unavailability of precise observations of these spectroscopic data, their accurate estimations are of great 
interest and useful in analyzing various astrophysical phenomena undergoing inside the heavenly bodies that contain As, Se and 
Br elements. An all-order perturbative many-body method in the relativistic theory framework has been employed to determine the atomic wave 
functions, which are further used to estimate the above quantities with the uncertainties. We found significant differences between some 
of our results and results that are available earlier.
\end{abstract}
\section{Introduction} \label{1}

White dwarfs, also known as degenerate dwarfs, are of great interest since early 1900s and were first discovered in triple star system of 40 
Edirani \citep{sutherland1972white}. They are typically highly densed stars having mass comparable to the Sun, and diameter 40 to 50 times smaller than the 
Sun \citep{burnham2013burnham}. The hot DO white dwarf, RE 0503-289, which lies within a degree of the hot DA white dwarf RE 0457-281 was detected 
in the ROSAT sky survey in 1994 \citep{Barstow}. The emission spectra of these white dwarfs showed abundance of photospheric lines of trans-iron 
group elements \citep{werner2012first}. Observation of spectral lines in the atmospheres of these high-gravity stars, e.g. spectral lines of
RE 0503-289 and G 191-B2B stars \citep{rauch2016stellar, werner2012first}, showed a reasonable abundance of As, Se and Br elements emitting 
ultraviolet (UV) spectra. Besides that the UV spectra obtained from the Far Ultraviolet Spectroscopic Explorer (FUSE), Goddard High Resolution 
Spectrograph (GHRS) on the Hubble Space Telescope and International Ultraviolet Explorer (IUE) for HD 149499 B and HZ 21 stars, which are cool 
DO white dwarfs, also showed the presence of As $(Z=33)$ and Se $(Z=34)$ and Br $(Z=35)$ spectral lines \citep{chayer2005abundance}. Atomic data 
are prerequisite for stellar-atmosphere modeling and the earlier available spectroscopic data of the above ions are less precise \citep{rauch2016stellar}.

Determination of chemical abundances and energy transport through the star depend upon the reliable values of oscillator strengths and 
transition probabilities of the spectral lines of the emission stars \citep{martin1992fine}. Accurate knowledge of these quantities are important
to analyze the intensities \citep{ruffoni2014fe} and infer fundamental parameters such as mass \textit{M}, radius \textit{R} and luminosity 
\textit{L} \citep{wittkowski2005fundamental} of the stellar objects. This very fact demands the need for more precise estimate of the  
spectroscopic properties of the emission lines of the elements present in the stellar objects. They are also useful in the analysis of 
interstellar and quasar absorption lines \citep{morton2000atomic}. In addition, the radiative data of the spectral lines of highly charged 
ions are essential for the construction of kinetic models of plasma processes and investigating processes in the thermonuclear reactor 
plasma \citep{glushkov1996calculation}. Presence of spectral lines of Cu-like ions motivates for more accurate determination 
of the radiative properties of these ions for analyzing the chemical abundances and inferring the stellar parameters that are essential for 
investigating the environmental conditions of the white dwarfs.

A considerable amount of relativistic data on oscillator strengths for some selected transitions of As V has already been available so far 
\citep{cheng1978energy,migdalek1979influence,curtis1989comprehensive,martin1992fine,lavin1994relativistic,engo1997comparison,owono2005core}, 
whereas only a limited data for the transitions in Se VI \citep{cheng1978energy,migdalek1979influence,curtis1989comprehensive,martin1992fine,
lavin1994relativistic,engo1997comparison,owono2005core} and Br VII \citep{cheng1978energy,migdalek1979influence,curtis1989comprehensive,
lavin1994relativistic,knystautas1977oscillator} are known in literature. In 1978, Cheng et~al. \citep{cheng1978energy} considered the As V, 
Se VI and Br VII ions for theoretical studies and reported their electric dipole (E1) transition properties between the low-lying states 
($n = 4,5,6$) using the Hartree-Fock method. Migdalek and Baylis \citep{migdalek1979influence} had studied the roles of electron correlations in 
the oscillator strengths of the $4S_{1/2} \rightarrow 4P_{1/2,3/2}$ and $4P_{1/2,3/2} \rightarrow 4D_{3/2,5/2}$ transitions in the Cu-isoelectronic 
As, Se and Br ions by considering the core-polarization effects in the relativistic theory framework. Victor and Taylor used semi-empirical model-potential 
methods for the calculation of absorption oscillator strengths for the $nS \rightarrow n'P$ transitions with $n,n'=4$ to $7$ in the Cu-like systems till 
$Z=42$ \citep{tables1983oscillator}. Curtis and Theodosiou had performed comprehensive calculations of {excitation energies, ionization potential,
E1 polarizabilities and lifetimes of the $4P$ and $4D$ states as well as the oscillator strengths for the $4S \rightarrow 4P_{1/2,3/2}, 
4P_{1/2,3/2} \rightarrow 4D_{3/2}$ and $4P_{3/2} \rightarrow 4D_{5/2}$ transitions} of Cu-isoelectronic sequence by combining the experimental 
data and matrix elements obtained using the Hartree-Slater potential to represent the ionic-core \citep{curtis1989comprehensive}. Martin et~al. 
calculated oscillator strengths for the D1 and D2 lines of the As V and Se VI ions by using Quantum Defect Orbital (QDO) and Relativistic Quantum 
Defect Orbital (RQDO) techniques \citep{martin1992fine}. Lavin et~al. computed oscillator strengths for fine structure transitions i.e. 
$4S \rightarrow 4P_{1/2,3/2}, 4P_{1/2,3/2} \rightarrow 4D_{3/2}$ and $4P_{3/2} \rightarrow 5S_{1/2}$ by employing the QDO and RQDO methods for 
Cu-like systems up to $Z=92$ \citep{lavin1994relativistic}. Sen and Puri had evaluated the E1 oscillator strengths corresponding to the Rydberg 
transition ($nS \rightarrow nP$) for As V and Se VI using local density functional including correlation effects \citep{sen1995slater}. Engo et~al. 
calculated oscillator strengths for the Cu-isoelectronic sequences using supersymmetry-inspired quantum-defect methods in their relativistic 
and quasi-relativistic formulations and compared their data with the then available literature \citep{engo1997comparison}. In 2005, Owono et~al.
evaluated oscillator strengths of the principal $4S \rightarrow 4P_{1/2,3/2}$ transitions in As V and Se VI ions by including explicitly 
core-polarization effects using the Dirac and quasi-relativistic quantum-defect radial wave functions \citep{owono2005core}. However, interest to 
theoretically investigate atomic properties of these ions date back to 1973 when G. Sorenson had calculated atomic lifetimes for the 
Cu-isoelectronic As and Se ions using foil-excitation method to study the systematic trends in f-values for the $4S \rightarrow 4P$ and 
$4P \rightarrow 4D$ transitions in As V and the $4P \rightarrow 4D$ transitions in Se VI \citep{sorensen1973atomic}. Fischer had calculated 
wavelength and oscillator strength for the $4S$--$4P$ transition for As V using to non-relativistic theory \citep{fischer1977oscillator}. 
Beam-foil spectroscopy was implemented for the calculation of radiative lifetimes and multiplet f-values for the $4P$, $4D$ and $5S$ terms for 
As V \citep{pinnington1981beam} and Se VI \citep{pinnington1981beam,bahr1982beam}. In case of Br VII, Andersen et~al. derived systematic trends 
of f-values from the lifetime measurements using the foil-exchange technique for the $nS$--$nP$ transitions \citep{andersen1973oscillator}. 
Knystautas and Drouin obtained oscillator strengths for the $4S_{1/2} \rightarrow 4P_{1/2,3/2}$ and $4P_{3/2} \rightarrow 4D_{5/2}$ transitions 
using beam-foil technique \citep{knystautas1977oscillator}. 

On the basis of evidence of spectral lines of Cu-isoelectronic sequence in white dwarfs, it is important to seek through atomic properties of 
these ions precisely. It has been observed from the aforementioned discussions that atomic properties of As V, Se VI and Br VII were
mostly calculated using the non-relativistic approaches or lower-order many-body methods such as quantum defect theory. These calculations 
show large differences among different results of the investigated spectroscopic properties. Thus, it is necessary to employ more 
advanced recently developed methods in the relativistic theory framework to determine spectroscopic properties of the above highly-stripped ions
that are immensely interesting for the astrophysical studies. Among the calculations carried out earlier, the most recent data calculated for 
these ions were obtained based on the radial wave functions determined using the quantum defect orbitals which included only the 
core-polarization effects \citep{owono2005core}. These screening effects are considered through an effective Hamiltonian used in the RQDO approach
and more valid in the core region of the ions \citep{charro1999oscillator}, but precise evaluation of radiative properties requires accurate 
behavior of the wave functions in the asymptotic region. To improve over these calculations, we have implemented here a relativistic all-order 
(AO) formalism for the computation of wave functions for the considered ions. Our AO approach accounts electron correlations among the core 
and valence electrons on an equal footing, hence it can describe wave functions properly in both the nuclear and asymptotic regions of an 
atomic system. It is also based on the fully four-component relativistic formalism, thereby it is adequate to incorporate the relativistic effects 
and to describe the fine-structure splitting of atomic levels in a natural manner. In the present study, we consider all the allowed transitions
among the $nS_{1/2}$, $nP_{1/2,3/2}$, $n'D_{3/2,5/2}$ and $n'F_{5/2,7/2}$ states with $n=4$ to $6$ and $n'=4,5$ for the above ions, and present 
wavelengths, line strengths, transition probabilities and absorption oscillator strengths of these transitions as well as the estimated lifetimes
of the excited states associated with these transitions. To validate the results, we have also calculated energies of a few low-lying states of these 
ions and compared them with the data listed in the National Institute of Standards and Technology Atomic Database (NIST AD) \citep{ralchenko2008nist}.

The paper is organized as follows: In Sec. \ref{2}, we provide theoretical formulae for different spectroscopic quantities such as transition probabilities, oscillator strengths and lifetimes of the atomic states, whereas in Sec. \ref{3} we discuss the method of 
atomic wave function evaluation. Sec. \ref{4} discusses all the acquired data from the present work and compare them with the previously reported 
values, while findings are concluded in Sec. \ref{5}. All the results are given in atomic units (a.u.) unless stated otherwise.

\section{Theoretical Aspects}
\label{2}

Transitions among different atomic states are generally driven through the E1 channel when allowed. The transition probability 
due to this channel ($A^{E1}_{vk}$) for the transition between the lower state $|\psi_{k}\rangle$ and the upper state $|\psi_{v}\rangle$ with
corresponding angular momenta $J_{k}$ and $J_{v}$ is given by \citep{kelleher2008atomic} 
\begin{equation}
A_{vk}^{E1}=\frac{2}{3} \alpha c \pi \sigma \times \left(\frac{\alpha \sigma}{R_{\infty}}\right)^2 \frac{S^{E1}}{g_{v}}, \label{eq2}
\end{equation}
where $\alpha= \frac{e^{2}}{2\epsilon_{0}hc}$ is the fine structure constant, $c$ is the speed of light, $g_v=2J_v+1$ is the degeneracy factor for the 
corresponding state, $R_{\infty}$ is the Rydberg constant, $S^{E1}$ denotes the line strength which can be calculated by the formula 
$S^{E1}=|\langle J_{v}||\textbf{D}|| J_{k} \rangle|^{2}$ \citep{nahar1995atomic} with the E1 operator $\textbf{D}=\Sigma_{j} \textbf{d}_{j}
=-e\Sigma_{j}\textbf{r}_{j}$ having $j^{th}$ electron at position $\textbf{r}_{j}$ and $\sigma$ denotes the differential energy between the  
transition levels given by $\sigma=E_{v}-E_{k}$. Substituting the values of fundamental constants, the absorption transition probability (in $s^{-1}$) 
is given by \citep{AYMAR1978537,kelleher2008atomic} 
\begin{equation}
A^{E1}_{vk}=\frac{2.02613 \times 10^{18}}{g_{v}\lambda^{3}}S^{E1}, \label{eq3}
\end{equation}
where $\lambda$ and $S$ are used in $\text{\normalfont\AA}$ and a.u., respectively.

Consequently, the oscillator strengths for the corresponding transitions are evaluated by using the following formulae \citep{AYMAR1978537,kelleher2008atomic}
\begin{equation}
f_{kv}^{E1}=\frac{1}{3\alpha}\left(\frac{\alpha\sigma}{R_{\infty}} \right) \times \frac{S^{E1}}{g_{k}}=\frac{303.756}{g_{k}\lambda}\times S^{E1}. \label{eq10}
\end{equation}

The radiative lifetime ($\tau$) of level $v$ can be estimated by taking the reciprocal of the total transition probability calculated by summing up the 
individual transition probabilities of the transitions from the considered upper electronic state ($v$) to each possible lower electronic 
state ($k$) \citep{qin2019energy}; i.e.
\begin{equation}
\tau_{v}=\frac{1}{\Sigma_{k} A^{E1}_{vk}}. \label{eq14}
\end{equation}
Substituting $A^{E1}_{vk}$ values from Eq. (\ref{eq3}), $\tau$ can be given in $s$.

\section{Method of Evaluation}

\label{3}

\subsection{Relativistic AO Method}

For the precise evaluation of transition-matrix elements, the electron correlation effects are included to all-order using our AO method which 
is based on the relativistic coupled-cluster (RCC) theory framework. The general formulation and potential applications of RCC theory, also 
referred as gold standard of many-body method, can be found in many earlier studies including Refs. \citep{blundell1991relativistic,
safronova2008all,sahoo2015correlation,sahoo2015theoretical}. We give a brief outline of our employed AO method in the RCC theory framework below. 

In the (R)CC theory ansatz, wave function of a many-electron system can be expressed as \citep{vcivzek1969use}
\begin{equation}
|\psi_0\rangle=e^S|\phi_0\rangle, \label{eqa}
\end{equation}
where $|\phi_0 \rangle$ is the mean-field wave function of an atomic state and $S$ represents the electron excitation operator from the mean-field
wave function. We have obtained the mean-field wave function using the Dirac-Fock (DF) method. First, we solve DF equation for the Ni-like 
closed-shell configurations for the undertaken ions then add a valence orbital to obtain the DF wave function of the Cu-like ions by defining 
as \citep{safronova1999relativistic}
\begin{equation}
|\phi_{v}\rangle=a_{v}^{\dagger}| \phi_0 \rangle . \label{eqdag}
\end{equation} 
Expanding $e^S$ in Eq. (\ref{eqa}), we get
\begin{equation}
|\psi_v\rangle=(1+S+\frac{S^2}{2}+... )|\phi_v\rangle . \label{eq17}
\end{equation}
For computational simplicity, we have dropped the non-linear terms and consider only the singly and doubly excited-state configurations (SD method)
in our AO approach by expressing \citep{blundell1989relativistic,safronova1998relativistic}
\begin{equation}
|\psi_v\rangle=(1+S_1+S_2)|\phi_v\rangle . \label{eq18}
\end{equation}
The excitation operators take into account excitations from both from the core and valence orbitals of the DF wave functions of the Cu-like ions, 
and they are defined using the second quantized operators as \citep{blundell1989relativistic,safronova1998relativistic,safronova1999relativistic,
iskrenova2007high}
\begin{equation}
S_1=\sum_{ma} \rho_{ma} a^{\dagger}_m a_a + \sum_{m\neq v}\rho_{mv} a^{\dagger}_m a_v, \label{eq19}
\end{equation}
and
\begin{equation}
S_2=\frac{1}{2}\sum_{mnab} \rho_{mnab} a^{\dagger}_m a^{\dagger}_n a_b a_a + \sum_{mna}\rho_{mnva} a^{\dagger}_m a^{\dagger}_n a_v a_a , \label{eq19}
\end{equation}
where the indices $m$ and $n$ range over all possible virtual states, the indices $a$ and $b$ range over all occupied core states and the 
indices $v$ and $w$ represent valence states of the system. The quantities $\rho_{ma}$ and $\rho_{mv}$ depict excitation coefficients of 
the respective singly-excitations for core and valence electron, whereas $\rho_{mnab}$ and $\rho_{mnva}$ denote doubly-excitation coefficients 
for core and valence electrons respectively. In addition to this, we have tried to improve the results by considering contributions from the 
dominant triple excitations by constructing triple excitation operator in the perturbative approach (SDpT) as 
\begin{eqnarray}\label{eq19}
 S_3^{\text{pert}} &=& \frac{1}{6}\sum_{mnrab}\rho_{mnrvab}a^{\dagger}_{m}a^{\dagger}_{n}a^{\dagger}_{r}a_{b}a_{a}a_{v}  \nonumber\\
&&  +\frac{1}{18}\sum_{mnrabc}\rho_{mnrabc}a^{\dagger}_{m}a^{\dagger}_{n}a^{\dagger}_{r}a_{c}a_{b}a_{a} ,
\end{eqnarray}
where $\rho_{mnrvab}$ and $\rho_{mnrabc}$ are triple excitation coefficients. This operator is considered in $S$ along with $S_1$ and $S_2$ 
to obtain amplitudes of the SD excitation operators as discussed in Refs. \citep{blundell1991relativistic,safronova1999relativistic,safronova2008all}.

After obtaining the wave function in both the SD and SDpT methods, the matrix elements for the E1 operator $D$ between states $v$ and $w$ with 
the corresponding wave functions $|\psi_{v}\rangle$ and $|\psi_{w}\rangle$ are evaluated by \citep{iskrenova2007high}
\begin{eqnarray}
D_{wv}=\frac{\langle\psi_{w}|D|\psi_{v}\rangle}{\sqrt{\langle\psi_{w}|\psi_{w}\rangle \langle\psi_{v}|\psi_{v}\rangle}}. \label{eq16}
\end{eqnarray}
In the resulting expression of the SD method, the numerator contains the sum of 20 terms for incorporating electron correlation effects
\citep{safronova2011blackbody} apart from the dominantly contributing DF expression.

\subsection{Gauge Invariance and Evaluation of Uncertainty}
\label{gi}

To ascertain numerical stability and convergence in the results for the determination of E1 matrix elements, we have calculated these quantities 
using the length (L) and velocity (V) gauge expressions of the E1 operator. Though the results from the exact RCC theory can be gauge independent, 
results from both the gauge expressions obtained using the approximated SD and SDpT methods can differ. Also, it is important to use the 
four-component Dirac orbitals for accurate calculations, which is given for an orbital $v$ with angular quantum numbers $j$ and $m_j$ as
\citep{sakurai2006advanced} 
\begin{equation}
|j_{v}m_{j_{v}} \rangle=\frac{1}{r} \left(\begin{matrix}
iG_v (r)| \chi_{\kappa_{v},m_{j_v}}(\theta,\phi) \rangle \\
F_v (r)|\chi_{-\kappa_{v},m_{j_v}}(\theta,\phi) \rangle \\
\end{matrix}\right),
\end{equation}
where $G_v (r)$ and $F_v (r)$ are the large and small components of Dirac wave function, respectively. Here, $|\chi_{\kappa_{v},m_{j_v}}
(\theta,\phi)\rangle $ is the angular function given by \citep{sakurai2006advanced}
\begin{equation}
\chi_{\kappa,m}(\theta,\phi)=\sum_{\sigma=\pm \frac{1}{2}} \langle lm-\sigma \frac{1}{2} \sigma|l \frac{1}{2}jm \rangle Y_{l}^{m-\sigma} 
(\theta,\phi) \phi_{\sigma},
\end{equation}
where $\langle lm-\sigma \frac{1}{2} \sigma|l \frac{1}{2}jm \rangle$ are Clebsch-Gordan coefficients, $Y_{l}^{m-\sigma}$ are normalized spherical
harmonics and $\phi_\sigma$ are two-component spinors with the values $\phi_{\frac{1}{2}}=\left(\begin{matrix}
1\\
0\\
\end{matrix} \right)$
and $\phi_{\frac{-1}{2}}=\left( \begin{matrix}
0\\
1\\
\end{matrix} \right).$
On the basis of this single particle orbital wave function, the reduced E1 matrix element between $|j_{v} m_{j_{v}} \rangle$ and 
$|j_{w} m_{j_{w}} \rangle$ is given by \citep{johnson2006lectures}
\begin{eqnarray}
&&\langle j_{v}||d_l||j_{w} \rangle=\frac{3}{k} \langle \kappa_{v} ||C_{1}||\kappa_{w} \rangle \nonumber \\
 &&\times \int_{0}^{\infty} dr \left\{j_1(kr)|G_{v}(r)G_{w}(r)+F_{v}(r)F_{w}(r)\right.
\nonumber\\
& & \left. + j_{2}(kr) \left[ \frac{\kappa_{v}-\kappa_{w}}{2}[G_{v}(r)F_{w}(r)+F_{v}(r)G_{w}(r)]\right. \right.
\nonumber\\
&& \left. \left. +[G_{v}(r)F_{w}(r)-F_{v}(r)G_{w}(r)]\right]\right\},
\nonumber\\
& &
\end{eqnarray}
in L-gauge and
\begin{eqnarray}
&& \langle j_v||d_v||j_w \rangle=\frac{3}{k} \langle \kappa_{v} ||C_{1}||\kappa_{w} \rangle \nonumber \\
 &&\times \int_{0}^{\infty} dr \left\{-\frac{\kappa_v - \kappa_w}{2}\left[\frac{dj_1(kr)}{dkr}+\frac{j_1(kr)}{kr}\right] \right.
\nonumber\\
& & \left. \times [G_{v}(r)F_{w}(r)+F_{v}(r)G_{w}(r)]\right.
\nonumber\\
&& \left. +\frac{j_1(kr)}{kr}[G_{v}(r)F_{w}(r)-F_{v}(r)G_{w}(r)]\right\},
\nonumber\\
& &
\end{eqnarray}
in V-gauge. Here, $k=\omega\alpha$ is the emitted photon energy $\omega$ of the transition, $C_n$ is the normalized spherical harmonic of rank $n$ and $j_l(x)$ is a spherical Bessel function of order $l$, given by
\begin{equation}
j_l(x)=\frac{x^l}{1\cdot{3}\cdot{5}\cdots(2l+1)}.
\end{equation} 

The differences in the results from both the gauge forms can be used as the maximum uncertainties corresponding to our calculated E1 matrix 
elements. Since calculations using L-gauge expression converge faster with respect to the number of configurations, we believe that these 
results are more reliable. Therefore, we consider these results as the central values for our further computations. We also analyze the
differences in the results from both the SD and SDpT methods to verify contributions from the neglected higher-level excitations.

\onecolumn
\begin{table}
\caption{\label{tab1} Our calculated energy values from the SD method ($E^{SD}$) in cm$^{-1}$ of a few low-lying states of the As V, Se VI and 
Br VII ions are listed. They are compared with the literature values and percentage (\%) of deviations ($\delta$) of our calculated values from
the experimental values from the NIST AD \citep{ralchenko2008nist} are also given. }
\begin{center}
\begin{tabular}{@{\extracolsep\fill}cccccccccc@{}}
\hline
State & \multicolumn{3}{c}{As V} & \multicolumn{3}{c}{Se VI} & \multicolumn{3}{c}{Br VII}\\
\cline{2-4}
\cline{5-7}
\cline{8-10}
 & $E^{SD}$ & Experiment & $\delta$ & $E^{SD}$ & Experiment & $\delta$ & $E^{SD}$ & Experiment & $\delta$ \\
\hline
$4S_{1/2}$ & $506608$ & $506250$ & $0.07$ & $660552$ & $659980$ & $0.10$ & $830186$ & $831000$ & $0.10$\\
$4P_{1/2}$ & $409410$ & $409115$ & $0.07$ & $547699$ & $547215$ & $0.09$ & $701807$ & $702726$ & $0.13$ \\
$4P_{3/2}$ & $405224$ & $405005$ & $0.05$ & $541898$ & $541518$ & $0.07$ & $694093$ & $695146$ & $0.15$\\
$4D_{3/2}$ & $269505$ & $269353$ & $0.06$ & $377425$ & $377150$ & $0.07$ & $501283$ & $503795$ & $0.50$\\
$4D_{5/2}$ & $269034$ & $268908$ & $0.05$ & $376708$ & $376471$ & $0.06$ & $500261$ & $502934$ & $0.53$\\
$5S_{1/2}$ & $242909$ & $242654$ & $0.10$ & $326788$ & $326386$ & $0.12$ & $420835$ & &\\
$4F_{5/2}$ & $174422$ & $174263$ & $0.09$ & $252169$ & & & $344617$ & &\\
$4F_{7/2}$ & $174433$ & $174175$ & $0.15$ & $252187$ & & & $344642$ & &\\
\hline
\end{tabular}
\end{center}
\end{table}

\begin{longtable}{@{\extracolsep\fill}ccrrrrrr@{}}
\caption{\label{tab2} The line strengths ($S_{vk}$) (in a.u.) calculated using both the L- and V-gauge expressions, wavelengths ($\lambda$) 
(in $\text{\normalfont\AA}$), transition probabilities ($A_{L_{vk}}$) in ($s^{-1}$) and absorption oscillator strengths ($f_{L_{kv}}$) for the As V ion
through E1 decay channel are presented in this table. Values in square brackets represent the order of 10. Uncertainties are given in parentheses.}\\
\hline
Upper State(v) & Lower State(k) & $\lambda$ (in $\text{\normalfont\AA}$) & \multicolumn{2}{c}{$S_{vk}$ (in a.u.)} & $A_{L_{vk}}$(in $s^{-1}$) & $f_{L_{kv}}$\\ [0.5ex]
 \cline{4-5} \\
& & & L & V & & \\
 \hline
$4P_{1/2}$ & $4S_{1/2}$ & $1028.835$ & $2.068[0]$ & $2.019[0]$ & $1.924(73)[9]$ & $3.053(78)[-1]$\\

$4P_{3/2}$ & $4S_{1/2}$ & $986.356$ & $4.153[0]$ & $4.040[0]$ & $2.192(89)[9]$ & $6.40(19)[-1]$\\

$4D_{3/2}$ & $4P_{1/2}$ & $714.768$ & $5.208[0]$ & $5.031[0]$  & $7.22(33)[9]$ & $1.107(39)[0]$\\
$4D_{3/2}$ & $4P_{3/2}$ & $736.813$ & $1.069[0]$ & $1.034[0]$ & $1.354(60)[9]$ & $1.102(38)[-1]$\\

{$4D_{5/2}$} & $4P_{3/2}$ & $734.264$ & $9.610[0]$ & $9.296[0]$  & $8.20(37)[9]$ & $9.94(34)[-1]$\\

$4F_{5/2}$ & $4D_{3/2}$ & $1051.720$ & $1.663[1]$ & $1.687[1]$ & $4.83(16)[9]$ & $1.201(21)[0]$\\
{$4F_{5/2}$} & $4D_{5/2}$ & $1056.957$ & $1.195[0]$ & $1.210[0]$ & {$3.42(11)[8]$} & {$5.722(93)[-2]$}\\
 
{$4F_{7/2}$} & $4D_{5/2}$ & $1057.074$ & $2.387[1]$ & $2.492[1]$ & $5.12(27)[9]$ & $1.143(51)[0]$\\

$5S_{1/2}$ & $4P_{1/2}$ & $600.595$ & $5.285[-1]$ & $5.098[-1]$ & $2.47(12)[9]$ & $1.337(50)[-1]$\\
$5S_{1/2}$ & $4P_{3/2}$ & $616.084$ & $1.153[0]$ & $1.109[0]$ & $5.00(25)[9]$ & $1.422(57)[-1]$\\

$5P_{1/2}$ & $4S_{1/2}$ & $335.378$ & $1.613[-2]$ & $1.588[-2]$ & $4.33(15)[8]$ & $7.30(14)[-3]$\\
$5P_{1/2}$ & $4D_{3/2}$ & $1637.506$ & $4.215[0]$ & $3.980[0]$ & $9.72(62)[8]$ & $1.96(11)[-1]$\\
$5P_{1/2}$ & $5S_{1/2}$ & $2900.867$ & $8.197[0]$ & $8.077[0]$ & $3.40(11)[8]$ & $4.292(76)[-1]$\\

$5P_{3/2}$ & $4S_{1/2}$ & $333.726$ & $2.280[-2]$ & $2.280[-2]$ & $3.107(93)[8]$ & $1.038(10)[-2]$\\
$5P_{3/2}$ & $4D_{3/2}$ & $1598.867$ & $8.136[-1]$ & $7.691[-1]$ & $1.008(64)[8]$ & $3.86(22)[-2]$\\
$5P_{3/2}$ & $4D_{5/2}$ & $1611.003$ & $7.420[0]$ & $7.012[0]$ & $8.99(57)[8]$ & $2.33(13)[-1]$\\
$5P_{3/2}$ & $5S_{1/2}$ & $2781.775$ & $1.634[1]$ & $1.608[1]$ & $3.84(13)[8]$ & $8.92(17)[-1]$\\

$5D_{3/2}$ & $4P_{1/2}$ & $442.136$ & $5.242[-1]$ & $1.225[-1]$ & $3.1(3.2)[9]$ & $1.80(1.9)[-1]$\\
$5D_{3/2}$ & $4P_{3/2}$ & $450.474$ & $1.156[-1]$ & $2.856[-2]$ & $6.4(6.4)[8]$ & $1.9(2.0)[-2]$\\
$5D_{3/2}$ & $4F_{5/2}$ & $11346.415$ & $1.990[1]$ & $1.061[1]$ & {$7(19)[6]$} & $9(24)[-2]$\\
$5D_{3/2}$ & $5P_{1/2}$ & $3968.171$ & $1.473[1]$ & $2.304[1]$ & $1.19(60)[8]$ & $5.6(2.8)[-1]$\\
$5D_{3/2}$ & $5P_{3/2}$ & $4215.014$ & $2.972[0]$ & $4.818[0]$ & $2.0(1.1)[7]$ & $5.4(2.9)[-2]$\\

$5D_{5/2}$ & $4P_{3/2}$ & $463.532$ & $1.311[0]$ & $3.832[-1]$ & $4.4(4.1)[9]$ & $2.1(2.0)[-1]$\\
$5D_{5/2}$ & $4F_{5/2}$ & $6636.920$ & $1.173[0]$ & $4.162[0]$ & $1.4(2.4)[6]$ & {$8.95(16)[-3]$}\\
$5D_{5/2}$ & $4F_{7/2}$ & $6641.526$ & $2.345[1]$ & $8.352[1]$ & $2.7(4.8)[7]$ & {$1.3(2.4)[-1]$}\\
$5D_{5/2}$ & $5P_{3/2}$ & $5723.824$ & $2.027[1]$ & $4.136[1]$ & $3.7(3.1)[7]$ & $2.7(2.3)[-1]$\\

$5F_{5/2}$ & $4D_{3/2}$ & $633.899$ & $3.956[-1]$ & $4.199[-1]$ & $5.25(35)[8]$ & $4.74(29)[-2]$\\
$5F_{5/2}$ & $4D_{5/2}$ & $635.797$ & $2.756[-2]$ & $2.936[-2]$ & $3.62(24)[7]$ & $2.19(13)[-3]$\\
$5F_{5/2}$ & $5D_{3/2}$ & $1398.897$ & $9.193[0]$ & $6.145[0]$ & $1.13(42)[9]$ & $5.0(1.8)[-1]$\\
$5F_{5/2}$ & $5D_{5/2}$ & $1286.359$ & $3.636[-1]$ & $2.190[-1]$ & $5.8(2.6)[7]$ & $1.43(64)[-2]$\\

{$5F_{7/2}$} & $4D_{5/2}$ & $635.780$ & $5.535[-1]$ & $5.868[-1]$ & $5.46(36)[8]$ & $4.41(26)[-2]$\\
$5F_{7/2}$ & $5D_{5/2}$ & $1286.285$ & $7.285[0]$ & $4.381[0]$ & $8.7(3.9)[8]$ & $2.9(1.3)[-1]$\\

$6S_{1/2}$ & $4P_{1/2}$ & $376.150$ & $5.244[-2]$ & $4.973[-2]$ & $9.98(60)[8]$ & $2.12(11)[-2]$\\
$6S_{1/2}$ & $4P_{3/2}$ & $382.168$ & $1.102[-1]$ & $1.043[-1]$ & $2.00(12)[9]$ & $2.19(12)[-2]$\\
$6S_{1/2}$ & $5P_{1/2}$ & $1541.372$ & $2.316[0]$ & $2.283[0]$ & $6.41(21)[8]$ & $2.283(40)[-1]$\\
$6S_{1/2}$ & $5P_{3/2}$ & $1577.251$ & $4.990[0]$ & $4.915[0]$ & $1.289(43)[9]$ & $2.403(44)[-1]$\\

$6P_{1/2}$ & $4S_{1/2}$ & $263.658$ & $6.241[-3]$ & $6.084[-3]$ & {$3.45(14)[8]$} & $3.595(98)[-3]$\\
$6P_{1/2}$ & $4D_{3/2}$ & $703.352$ & $1.069[-1]$ & $9.610[-2]$ & $3.11(34)[8]$ & $1.15(12)[-2]$\\
$6P_{1/2}$ & $5S_{1/2}$ & $865.200$ & $2.890[-2]$ & $2.993[-2]$ & $4.52(21)[7]$ & $5.07(19)[-3]$\\\
$6P_{1/2}$ & $5D_{3/2}$ & $1788.677$ & $2.274[0]$ & $2.993[0]$ & $4.0(1.2)[8]$ & $9.7(2.8)[-2]$\\
$6P_{1/2}$ & $6S_{1/2}$ & $6161.241$ & $2.231[1]$ & $2.209[1]$ & $9.66(30)[7]$ & $5.499(77)[-1]$\\

$6P_{3/2}$ & $4S_{1/2}$ & $263.170$ & $9.801[-3]$ & $9.409[-3]$ & $2.72(14)[8]$ & $5.66(24)[-3]$\\
$6P_{3/2}$ & $4D_{3/2}$ & $699.886$ & $2.220[-2]$ & $2.016[-2]$ & $3.28(32)[7]$ & $2.41(23)[-3]$\\
{$6P_{3/2}$} & $4D_{5/2}$ & $702.202$ & $1.998[-1]$ & $1.798[-1]$ & $2.92(31)[8]$ & {$1.44(15)[-2]$}\\
$6P_{3/2}$ & $5S_{1/2}$ & $859.961$ & $3.648[-2]$ & $3.803[-2]$ & $2.91(15)[7]$ & $6.44(28)[-3]$\\
$6P_{3/2}$ & $5D_{3/2}$ & $1766.430$ & $4.225[-1]$ & $4.316[-1]$ & $3.9(1.2)[7]$ & $1.82(57)[-2]$\\
$6P_{3/2}$ & $5D_{5/2}$ & $1590.705$ & $1.858[0]$ & $2.699[0]$ & $2.34(96)[8]$ & {$5.9(2.4)[-2]$}\\
$6P_{3/2}$ & $6S_{1/2}$ & $5905.068$ & $4.433[1]$ & $4.436[1]$ & $1.090(33)[8]$ & $1.140(11)[0]$\\

\hline
\end{longtable}

\begin{longtable}{@{\extracolsep\fill}ccrrrrrr@{}}
\caption{\label{tab3} The line strengths ($S_{vk}$) (in a.u.) calculated using both the L and V-gauge expressions, wavelengths ($\lambda$) 
(in $\text{\normalfont\AA}$), transition probabilities ($A_{L_{vk}}$) in ($s^{-1}$) and absorption oscillator strengths ($f_{L_{kv}}$) for the Se VI ion
through E1 decay channel are presented in this table. Values in square brackets represent the order of 10. Uncertainties are given in 
parentheses. }\\
\hline
Upper State(v) & Lower State(k) & $\lambda$ (in $\text{\normalfont\AA}$) & \multicolumn{2}{c}{$S_{vk}$ (in a.u.)} & $A_{L_{vk}}$(in $s^{-1}$) & $f_{L_{kv}}$\\ [0.5ex]
 & & & L & V & & \\
\hline
$4P_{1/2}$ & $4S_{1/2}$ & $886.111$ & $1.484[0]$ & $1.486[0]$ & $2.160(65)[9]$ & $2.543(26)[-1]$\\

$4P_{3/2}$ & $4S_{1/2}$ & $842.783$ & $2.983[0]$ & $2.989[0]$ & $2.524(76)[9]$ & $5.375(55)[-1]$\\

$4D_{3/2}$ & $4P_{1/2}$ & $587.287$ & $3.806[0]$ & $3.810[0]$ & $9.52(29)[9]$ & $9.844(99)[-1]$\\
$4D_{3/2}$ & $4P_{3/2}$ & $608.003$ & $7.815[-1]$ & $7.832[-1]$ & $1.761(53)[9]$ & $9.76(10)[-2]$\\

$4D_{5/2}$ & $4P_{3/2}$ & $605.366$ & $7.033[0]$ & $7.038[0]$ & $1.071(32)[10]$ & $8.823(88)[-1]$\\

$4F_{5/2}$ & $4D_{3/2}$ & $798.368$ & $1.178[1]$ & $1.179[1]$ & {$7.82(23)[9]$} & $1.120(11)[0]$\\
$4F_{5/2}$ & $4D_{5/2}$ & $802.961$ & $8.446[-1]$ & $8.464[-1]$ & $5.51(17)[8]$ & $5.235(54)[-2]$\\
 
$4F_{7/2}$ & $4D_{5/2}$ & $803.074$ & $1.691[1]$ & $1.692[1]$ & $8.27(25)[9]$ & $1.066(11)[0]$\\

$5S_{1/2}$ & $4P_{1/2}$ & $452.671$ & $4.007[-1]$ & $3.994[-1]$ & $4.38(13)[9]$ & $1.344(14)[-1]$\\
$5S_{1/2}$ & $4P_{3/2}$ & $464.880$ & $8.780[-1]$ & $8.761[-1]$ & $8.85(27)[9]$ & $1.434(15)[-1]$\\

$5P_{1/2}$ & $4S_{1/2}$ & $266.267$ & $2.924[-2]$ & $2.890[-2]$ & $1.569(51)[9]$ & $1.668(26)[-2]$\\
$5P_{1/2}$ & $4D_{3/2}$ & $1081.842$ & $2.746[0]$ & $2.739[0]$ & $2.197(66)[9]$ & $1.927(20)[-1]$\\
$5P_{1/2}$ & $5S_{1/2}$ & $2392.438$ & $6.295[0]$ & $6.295[0]$ & $4.66(14)[8]$ & $3.996(40)[-1]$\\

$5P_{3/2}$ & $4S_{1/2}$ & $264.773$ & $4.622[-2]$ & $4.580[-2]$ & $1.261(40)[9]$ & $2.652(36)[-2]$\\
$5P_{3/2}$ & $4D_{3/2}$ & $1057.594$ & $5.271[-1]$ & $5.256[-1]$ & $2.257(68)[8]$ & $3.785(39)[-2]$\\
$5P_{3/2}$ & $4D_{5/2}$ & $1065.669$ & $4.814[0]$ & $4.800[0]$ & $2.015(61)[9]$ & $2.287(24)[-1]$\\
$5P_{3/2}$ & $5S_{1/2}$ & $2276.988$ & $1.257[1]$ & $1.256[1]$ & $5.39(16)[8]$ & $8.382(84)[-1]$\\

$5D_{3/2}$ & $4P_{1/2}$ & $314.540$ & $1.369[-3]$ & $1.296[-3]$ & $2.23(14)[7]$ & $6.61(36)[-4]$\\
$5D_{3/2}$ & $4P_{3/2}$ & $320.387$ & $2.250[-4]$ & $2.250[-4]$ & $3.47(10)[6]$ & $5.333(53)[-5]$\\
$5D_{3/2}$ & $4F_{5/2}$ & $4465.359$ & $7.344[-2]$ & $9.120[-2]$ & $4.18(96)[5]$ & {$8.3(1.9)[-4]$}\\
$5D_{3/2}$ & $5P_{1/2}$ & $1811.094$ & $1.232[-1]$ & $1.225[-1]$ & $1.051(32)[7]$ & $1.033(12)[-2]$\\
$5D_{3/2}$ & $5P_{3/2}$ & $1883.383$ & $2.528[-2]$ & $2.528[-2]$ & $1.917(58)[6]$ & $1.019(10)[-3]$\\

$5D_{5/2}$ & $4P_{3/2}$ & $328.381$ & $1.521[-3]$ & $1.369[-3]$ & $1.45(16)[7]$ & $3.52(36)[-4]$\\
$5D_{5/2}$ & $4F_{5/2}$ & $6758.432$ & $2.209[-3]$ & $3.481[-3]$ & $2.4(1.2)[3]$ & {$1.66(85)[-5]$}\\
$5D_{5/2}$ & $4F_{7/2}$ & $6750.409$ & $4.410[-2]$ & $7.182[-2]$ & $4.8(2.7)[4]$ & {$2.5(1.4)[-4]$}\\
$5D_{5/2}$ & $5P_{3/2}$ & $2197.916$ & $1.043[-1]$ & $1.030[-1]$ & $3.32(11)[6]$ & $3.605(57)[-3]$\\

$5F_{5/2}$ & $4D_{3/2}$ & $463.324$ & $1.076[-1]$ & $1.082[-1]$ & $3.65(11)[8]$ & $1.763(21)[-2]$\\
$5F_{5/2}$ & $4D_{5/2}$ & $464.867$ & $7.396[-3]$ & $7.396[-3]$ & $2.486(75)[7]$ & $8.055(81)[-4]$\\
$5F_{5/2}$ & $5D_{3/2}$ & $1466.674$ & $2.352[-1]$ & $2.480[-1]$ & $2.52(15)[7]$ & $1.218(66)[-2]$\\
$5F_{5/2}$ & $5D_{5/2}$ & $1319.613$ & $7.225[-3]$ & $7.744[-3]$ & $1.062(81)[6]$ & {$2.77(20)[-4]$}\\

$5F_{7/2}$ & $4D_{5/2}$ & $464.886$ & $1.459[-1]$ & $1.467[-1]$ & $3.68(11)[8]$ & $1.589(18)[-2]$\\
$5F_{7/2}$ & $5D_{5/2}$ & $1319.765$ & $1.391[-1]$ & $1.498[-1]$ & $1.53(12)[7]$ & {$5.34(40)[-3]$}\\

$6S_{1/2}$ & $4P_{1/2}$ & $284.481$ & $4.244[-2]$ & $4.203[-2]$ & $1.867(59)[9]$ & $2.266(32)[-2]$\\
$6S_{1/2}$ & $4P_{3/2}$ & $289.255$ & $8.940[-2]$ & $8.880[-2]$ & $3.74(12)[9]$ & $2.347(28)[-2]$\\
$6S_{1/2}$ & $5P_{1/2}$ & $1126.022$ & $1.638[0]$ & $1.636[0]$ & $1.163(35)[9]$ & $2.210(22)[-1]$\\
$6S_{1/2}$ & $5P_{3/2}$ & $1153.550$ & $3.557[0]$ & $3.553[0]$ & $2.348(70)[9]$ & $2.342(24)[-1]$\\

$6P_{1/2}$ & $4S_{1/2}$ & $206.419$ & $1.082[-2]$ & $1.061[-2]$ & $1.246(44)[9]$ & $7.96(17)[-3]$\\
$6P_{1/2}$ & $4D_{3/2}$ & $496.711$ & $9.303[-2]$ & $9.303[-2]$ & $7.69(23)[8]$ & $1.422(14)[-2]$\\
$6P_{1/2}$ & $5S_{1/2}$ & $663.624$ & $4.666[-2]$ & $4.623[-2]$ & $1.617(51)[8]$ & $1.068(15)[-2]$\\\
$6P_{1/2}$ & $5D_{3/2}$ & $1863.092$ & $4.080[-2]$ & $4.121[-2]$ & $6.39(20)[6]$ & $1.663(23)[-3]$\\
$6P_{1/2}$ & $6S_{1/2}$ & $4979.779$ & $1.711[1]$ & $1.709[1]$ & $1.403(42)[8]$ & $5.217(52)[-1]$\\

$6P_{3/2}$ & $4S_{1/2}$ & $205.979$ & $1.823[-2]$ & $1.823[-2]$ & $1.056(32)[9]$ & $1.344(13)[-2]$\\
$6P_{3/2}$ & $4D_{3/2}$ & $494.170$ & $1.932[-2]$ & $1.904[-2]$ & $8.11(27)[7]$ & $2.969(52)[-3]$\\
$6P_{3/2}$ & $4D_{5/2}$ & $495.926$ & $1.722[-1]$ & $1.722[-1]$ & $7.15(21)[8]$ & {$1.758(18)[-2]$}\\
$6P_{3/2}$ & $5S_{1/2}$ & $659.096$ & $6.554[-2]$ & $6.554[-2]$ & $1.159(35)[8]$ & $1.510(15)[-2]$\\
$6P_{3/2}$ & $5D_{3/2}$ & $1827.838$ & $7.569[-3]$ & $8.100[-3]$ & $6.28(19)[5]$ & $3.145(31)[-4]$\\
$6P_{3/2}$ & $5D_{5/2}$ & $1604.937$ & $1.664[-2]$ & $1.690[-2]$ & $2.039(69)[6]$ & {$5.249(97)[-4]$}\\
$6P_{3/2}$ & $6S_{1/2}$ & $4735.642$ & $3.402[1]$ & $3.394[1]$ & $1.623(49)[8]$ & $1.091(11)[0]$\\

\hline\\
\end{longtable}

\begin{longtable}{@{\extracolsep\fill}ccrrrrrr@{}}
\caption{\label{tab4} The line strengths ($S_{vk}$) (in a.u.) calculated using both the L and V-gauge expressions, wavelengths ($\lambda$) 
(in $\text{\normalfont\AA}$), transition probabilities ($A_{L_{vk}}$) in ($s^{-1}$) and absorption oscillator strengths ($f_{L_{kv}}$) for the Br VII ion
through E1 decay channel are presented in this table. Values in square brackets represent the order of 10. Uncertainties are given in parentheses.}\\
\hline
Upper State(v) & Lower State(k) & $\lambda$ (in $\text{\normalfont\AA}$) & \multicolumn{2}{c}{$S_{vk}$ (in a.u.)} & $A_{L_{vk}}$(in $s^{-1}$) & $f_{L_{kv}}$\\ [0.5ex]
 & & & L & V & & \\
 \hline
$4P_{1/2}$ & $4S_{1/2}$ & $778.944$ & $1.272[0]$ & $1.275[0]$ & {$2.727(82)[9]$} & $2.481(25)[-1]$\\

$4P_{3/2}$ & $4S_{1/2}$  & $734.789$ & $2.560[0]$ & $2.563[0]$ & {$3.269(98)[9]$} & $5.291(53)[-1]$\\

$4D_{3/2}$ & $4P_{1/2}$ & $498.692$ & $3.183[0]$ & $3.183[0]$ & $1.300(39)[10]$ & $9.693(97)[-1]$\\
$4D_{3/2}$ & $4P_{3/2}$ & $518.646$ & $6.529[-1]$ & $6.545[-1]$ & {$2.370(71)[9]$} & $9.559(98)[-2]$\\

$4D_{5/2}$ & $4P_{3/2}$ & $515.912$ & $5.876[0]$ & $5.881[0]$ & $1.445(43)[10]$ & $8.649(87)[-1]$\\

$4F_{5/2}$ & $4D_{3/2}$ & $638.301$ & $9.145[0]$ & $9.151[0]$ & $1.187(36)[10]$ & $1.088(11)[0]$\\
$4F_{5/2}$ & $4D_{5/2}$ & $642.491$ & $6.561[-1]$ & $6.651[-1]$ & $8.35(25)[8]$ & $5.170(52)[-2]$\\
 
$4F_{7/2}$ & $4D_{5/2}$ & $642.593$ & $1.313[1]$ & $1.314[1]$ & $1.254(38)[10]$ & {$1.035(10)[0]$}\\

$5S_{1/2}$ & $4P_{1/2}$ & $355.906$ & $3.080[-1]$ & $3.069[-1]$ & $6.92(21)[9]$ & $1.314(14)[-1]$\\
$5S_{1/2}$ & $4P_{3/2}$ & $365.954$ & $6.806[-1]$ & $6.773[-1]$ & $1.407(43)[10]$ & $1.412(16   )[-1]$\\

$5P_{1/2}$ & $4S_{1/2}$ & $218.080$ & $3.028[-2]$ & $2.993[-2]$ & $2.957(95)[9]$ & $2.109(32)[-2]$\\
$5P_{1/2}$ & $4D_{3/2}$ & $771.345$ & $1.904[0]$ & $1.899[0]$ & $4.20(13)[9]$ & $1.875(20)[-1]$\\
$5P_{1/2}$ & $5S_{1/2}$ & $2032.719$ & $5.157[0]$ & $5.157[0]$ & $6.22(19)[8]$ & $3.853(39)[-1]$\\

$5P_{3/2}$ & $4S_{1/2}$ & $216.709$ & $4.796[-2]$ & $4.752[-2]$ & $2.387(75)[9]$ & $3.361(46)[-2]$\\
$5P_{3/2}$ & $4D_{3/2}$ & $754.465$ & $3.624[-1]$ & $3.612[-1]$ & $4.27(13)[8]$ & $3.648(38)[-2]$\\
$5P_{3/2}$ & $4D_{5/2}$ & $760.327$ & $3.316[0]$ & $3.309[0]$ & $3.82(12)[9]$ & {$2.208(23)[-1]$}\\
$5P_{3/2}$ & $5S_{1/2}$ & $1919.541$ & $1.030[1]$ & $1.030[1]$ & $7.38(22)[8]$ & $8.153(82)[-1]$\\

$5D_{3/2}$ & $4P_{1/2}$ & $243.347$ & $8.410[-4]$ & $8.410[-4]$ & $2.956(9)[7]$ & $5.249(52)[-4]$\\
$5D_{3/2}$ & $4P_{3/2}$ & $248.002$ & $4.000[-6]$ & $4.000[-6]$ & $1.328(40)[5]$ & $1.225(12)[-6]$\\
$5D_{3/2}$ & $4F_{5/2}$ & $1860.601$ & $6.111[0]$ & $6.101[0]$ & $4.81(14)[8]$ & $1.663(69)[-1]$\\
$5D_{3/2}$ & $5P_{1/2}$ & $1238.108$ & $1.116[1]$ & $1.116[1]$ & $2.979(89)[9]$ & $1.369(14)[0]$\\
$5D_{3/2}$ & $5P_{3/2}$ & $1284.228$ & $2.295[0]$ & $2.295[0]$ & $5.49(17)[8]$ & $1.357(14)[-1]$\\

$5D_{5/2}$ & $4P_{3/2}$ & $248.154$ & $7.290[-4]$ & $7.290[-4]$ & {$1.61(48)[7]$} & $2.231(22)[-4]$\\
$5D_{5/2}$ & $4F_{5/2}$ & $1869.183$ & $2.852[-1]$ & $2.862[-1]$ & $1.474(45)[7]$ & {$7.723(83)[-3]$}\\
$5D_{5/2}$ & $4F_{7/2}$ & $1868.321$ & $5.698[0]$ & $5.731[0]$ & $2.950(90)[8]$ & {$1.158(13)[-1]$}\\
$5D_{5/2}$ & $5P_{3/2}$ & $1288.311$ & $1.370[1]$ & $1.368[1]$ & $2.164(65)[9]$ & $8.078(82)[-1]$\\

$5F_{5/2}$ & $4D_{3/2}$ & $356.635$ & $2.102[-2]$ & $2.132[-2]$ & $1.565(52)[8]$ & $4.477(76)[-3]$\\
$5F_{5/2}$ & $4D_{5/2}$ & $357.939$ & $1.296[-3]$ & $1.369[-3]$ & $9.54(60)[6]$ & {$1.83(10)[-4]$}\\
$5F_{5/2}$ & $5D_{3/2}$ & $1428.839$ & $2.723[1]$ & $2.724[1]$ & $3.152(95)[9]$ & $1.447(14)[0]$\\
$5F_{5/2}$ & $5D_{5/2}$ & $1423.818$ & $1.290[0]$ & $1.295[0]$ & $1.510(46)[8]$ & {$4.588(49)[-2]$}\\

$5F_{7/2}$ & $4D_{5/2}$ & $357.958$ & $2.624[-2]$ & $2.657[-2]$ & $1.449(47)[8]$ & $3.71(41)[-3]$\\
$5F_{7/2}$ & $5D_{5/2}$ & $1424.119$ & $2.582[1]$ & $2.591[1]$ & $2.264(68)[9]$ & {$9.178(97)[-1]$}\\

$6S_{1/2}$ & $4P_{1/2}$ & $224.204$ & $3.349[-2]$ & $3.349[-2]$ & $3.010(90)[9]$ & $2.269(23)[-2]$\\
$6S_{1/2}$ & $4P_{3/2}$ & $228.150$ & $7.129[-2]$ & $7.129[-2]$ & $6.08(18)[9]$ & $2.373(24)[-2]$\\
$6S_{1/2}$ & $5P_{1/2}$ & $863.238$ & $1.210[0]$ & $1.208[0]$ & $1.906(57)[9]$ & {$2.129(22)[-1]$}\\
$6S_{1/2}$ & $5P_{3/2}$ & $885.407$ & $2.650[0]$ & $2.647[0]$ & $3.87(12)[9]$ & $2.273(23)[-1]$\\

$6P_{1/2}$ & $4S_{1/2}$ & $167.098$ & $1.020[-2]$ & $1.020[-2]$ & $2.215(66)[9]$ & $9.272(93)[-3]$\\
$6P_{1/2}$ & $4D_{3/2}$ & $370.990$ & $7.840[-2]$ & $7.784[-2]$ & $1.555(48)[9]$ & $1.605(20)[-2]$\\
$6P_{1/2}$ & $5S_{1/2}$ & $528.819$ & $5.244[-2]$ & $5.244[-2]$ & $3.59(11)[8]$ & $1.506(15)[-2]$\\\
$6P_{1/2}$ & $5D_{3/2}$ & $1690.989$ & $7.209[0]$ & $7.193[0]$ & $1.510(45)[9]$ & $3.238(33)[-1]$\\
$6P_{1/2}$ & $6S_{1/2}$ & $4157.961$ & $1.376[1]$ & $1.373[1]$ & $1.940(58)[8]$ & $5.030(51)[-1]$\\

$6P_{3/2}$ & $4S_{1/2}$ & $166.701$ & $1.716[-2]$ & $1.716[-2]$ & $1.876(56)[9]$ & $1.564(16)[-2]$\\
$6P_{3/2}$ & $4D_{3/2}$ & $369.040$ & $1.588[-2]$ & $1.588[-2]$ & $1.600(48)[8]$ & $3.267(33)[-3]$\\
$6P_{3/2}$ & $4D_{5/2}$ & $370.436$ & $1.436[-1]$ & $1.436[-1]$ & $1.431(44)[9]$ & {$1.963(22)[-2]$}\\
$6P_{3/2}$ & $5S_{1/2}$ & $524.865$ & $7.563[-2]$ & $7.563[-2]$ & $2.649(79)[8]$ & $2.188(22)[-2]$\\
$6P_{3/2}$ & $5D_{3/2}$ & $1651.209$ & $1.376[0]$ & $1.374[0]$ & $1.548(47)[8]$ & $6.328(64)[-2]$\\
$6P_{3/2}$ & $5D_{5/2}$ & $1644.508$ & $8.037[0]$ & $8.020[0]$ & $9.15(28)[8]$ & {$2.474(25)[-1]$}\\
$6P_{3/2}$ & $6S_{1/2}$ & $3925.430$ & $2.739[1]$ & $2.731[1]$ & $2.294(69)[8]$ & $1.060(11)[0]$\\

\hline
\end{longtable}

\begin{longtable}{@{\extracolsep\fill}ccrrrrrr@{}}
\caption{\label{tab6} Comparison of oscillator strengths for the As V, Se VI and Br VII ions from our calculations with available theoretical 
data and experimental values obtained using the beam-foil measurement.}\\
\hline
Ion & Transition & \multicolumn{5}{c}{Theoretical studies} & Experiment \\
\cline{3-7 }\\
 &  & Present & RMP-CP$^{a}$ & MCDF$^{b}$ & RQDO$^{c,d}$ & DSQDT$^{e,f}$ & Beam-foil$^g$\\
\hline
As V & $4S_{1/2} \rightarrow 4P_{1/2}$ & $0.305(78)$ & $0.254$ & $0.26159$ & $0.2793^{c}$ & $0.2571^e$,$0.2571^f$ &\\ 
& $4S_{1/2} \rightarrow 4P_{3/2}$ & $0.6395(19)$ & $0.531$ & $0.54797$ & $0.5779^{c}$ & $0.5453^{e}$,$0.5455^f$ &\\
& $4S_{1/2} \rightarrow 5P_{1/2}$ & $0.00730(14)$ & & & & $0.0054^{e}$ &\\
& $4S_{1/2} \rightarrow 5P_{3/2}$ & $0.01038(10)$ & & & & $0.0073^{e}$ &\\
& $4P_{1/2} \rightarrow 4D_{3/2}$ & $1.107(39)$ & $0.973$ & $0.98927$ & $0.8418^{d}$ & &\\
& $4P_{3/2} \rightarrow 4D_{3/2}$ & $0.1102(38)$ & $0.0970$ & $0.09869$ & $0.08479^{d}$ & &\\
& $4P_{3/2} \rightarrow 4D_{5/2}$ & $0.994(34)$ & $0.864$ & $0.89009$ & $0.7628^{d}$ & &\\
& $4P_{1/2} \rightarrow 5S_{1/2}$ & $0.1337(50)$ & & & $0.1266^{d}$ & &\\
& $4P_{3/2} \rightarrow 5S_{1/2}$ & $0.1422(57)$ & & & $0.1308^{d}$ & &\\
& $4P_{1/2} \rightarrow 6S_{1/2}$ & $0.0212(11)$ & & & $0.01843^{c}$ & &\\
& $4P_{3/2} \rightarrow 6S_{1/2}$ & $0.0219(12)$ & & & $0.01843^{c}$ & &\\
& $5P_{1/2} \rightarrow 6S_{1/2}$ & $0.228(40)$ & & & $0.2407^{c}$ & &\\
& $5P_{3/2} \rightarrow 6S_{1/2}$ & $0.240(44)$ & & & $0.2484^{c}$ & &\\
Se VI & $4S_{1/2} \rightarrow 4P_{1/2}$ & $0.2543(26)$ & $0.247$ & $0.2560$ & $0.2748^d$ & $0.2562^e$,$0.2562^f$ &\\
& $4S_{1/2} \rightarrow 4P_{3/2}$ & $0.5375(55)$ & $0.521$ & $0.5409$ & $0.5736^d$ & $0.5465^e$,$0.5465^f$ &\\
& $4S_{1/2} \rightarrow 5P_{1/2}$ & $0.01668((26)$ & & & & $0.0086^{e}$ &\\
& $4S_{1/2} \rightarrow 5P_{3/2}$ & $0.02652(36)$ & & & & $0.0126^{e}$ &\\
& $4P_{1/2} \rightarrow 4D_{3/2}$ & $0.9844(99)$ & $0.963$ & $1.0087$ & $0.8713^d$ & &\\
& $4P_{3/2} \rightarrow 4D_{3/2}$ & $0.09760(10)$ & $0.0959$ & $0.1001$ & $0.08725^d$ & &\\
& $4P_{3/2} \rightarrow 4D_{5/2}$ & $0.8823(88)$ & $0.864$ & $0.9035$ & $0.7859^d$ & &\\
& $4P_{1/2} \rightarrow 5S_{1/2}$ & $0.1344(14)$ & & & $0.1221^d$ & &\\
& $4P_{3/2} \rightarrow 5S_{1/2}$ & $0.1434(15)$ & & & $0.1266^d$ & &\\
& $4P_{1/2} \rightarrow 6S_{1/2}$ & $0.02266(32)$ & & & $0.01837^c$ && \\
& $4P_{3/2} \rightarrow 6S_{1/2}$ & $0.02347(28)$ & & & $0.01848^c$ & &\\
& $5P_{1/2} \rightarrow 6S_{1/2}$ & $0.2210(22)$ & & & $0.2298^c$ & &\\
& $5P_{3/2} \rightarrow 6S_{1/2}$ & $0.2342(24)$ & & & $0.2385^c$ & &\\
Br VII & $4S_{1/2} \rightarrow 4P_{1/2}$ & $0.2481(25)$ & $0.244$ & $0.2497$ & $0.2683^d$ & & $0.16$\\
& $4S_{1/2} \rightarrow 4P_{3/2}$ & $0.5291(53)$ & $0.521$ & $0.5321$ & $0.5652^d$ & & $0.32$\\
& $4P_{1/2} \rightarrow 4D_{3/2}$ & $0.9693(97)$ & $0.951$ & $0.9685$ & $0.8797^d$ & &\\
& $4P_{3/2} \rightarrow 4D_{3/2}$ & $0.09559(98)$ & $0.0938$ & $0.0956$ & $0.08761^d$ & &\\
& $4P_{3/2} \rightarrow 4D_{5/2}$ & $0.8649(87)$ & $0.848$ & $0.8643$ & $0.7901^d$ & & $0.34$\\
& $4P_{1/2} \rightarrow 5S_{1/2}$ & $0.1314(14)$ & & & $0.1208^d$ & &\\
& $4P_{3/2} \rightarrow 5S_{1/2}$ & $0.1412(16)$ & & & $0.1257^d$ & & \\
\hline
\end{longtable}
$^{a}$\citep{migdalek1979influence}\\
$^{b}$\citep{curtis1989comprehensive}\\
$^{c}$\citep{martin1992fine}\\
$^{d}$\citep{lavin1994relativistic}\\ 
$^{e}$\citep{engo1997comparison}\\ 
$^{f}$\citep{owono2005core}\\ 
$^g$\citep{knystautas1977oscillator}\\
\clearpage
\begin{longtable}{@{\extracolsep\fill}crrr@{}}
\caption{\label{tab5} The estimated lifetimes $\tau$ (in ns) for a few low-lying excited states of the As V, Se VI and Br VII ions and their 
comparisons with the available literature data. Uncertainties are given in the parentheses.}\\
\hline
State & As V & Se VI & Br VII\\ 
\hline
$4P_{1/2}$ & $0.520(20)$ & $0.4630(14)$ & $0.3667(11)$\\
 & $0.68 \pm{0.09}^a$ & $0.45 \pm{0.05}^{a,c}$ & \\
 & $0.607^b$ & $0.460^b$ & $0.365^b$\\
$4P_{3/2}$ & $0.456(19)$ & $0.3962(12)$ & $0.3059(9)$\\
 & $0.54 \pm{0.03}^a$ & $0.39 \pm{0.04}^{a,c}$ & \\
  & $0.534^b$ & $0.395^b$ & $0.305^b$\\
$4D_{3/2}$ & $0.117(5)$ & $0.0887(23)$ & $0.0651(17)$\\ 
 & $0.166 \pm{0.012}^a$ & $0.11 \pm{0.02}^{a,c}$ & \\
 & $0.131^b$ & ${0.087}^b$ & {$0.065^b$}\\
$4D_{5/2}$ & $0.122(6)$ & $0.0934(28)$ & $0.0692(21)$\\
 & $0.18\pm{0.02}^a$ & $0.14 \pm{0.03}^{a,c}$ & \\
 & {$0.136^b$} & {$0.091^b$} & {$0.069^b$}\\
 $5S_{1/2}$ & $0.134(5)$ & $0.0756(17)$ & $0.0476(11)$\\
 & $0.14 \pm{0.01}^a$ & $0.06 \pm{0.02}^{a,c}$ & \\
$4F_{5/2}$ & $0.194(6)$ & $0.1195(33)$ & $0.0787(22)$\\
$4F_{7/2}$ & $0.195(10)$ & $0.1209(36)$ & $0.0797(24)$\\
$5P_{1/2}$ & $0.573(21)$ & $0.2363(47)$ & $0.1285(27)$\\
$5P_{3/2}$ & $0.589(21)$ & $0.2475(46)$ & $0.1356(26)$\\
$5D_{3/2}$ & $0.26(22)$ & $25.90(96)$ & $0.2476(56)$\\
$5D_{5/2}$ & $0.22(20)$ & $55.96(5.02)$ & {$0.4016(11)$}\\
$5F_{5/2}$ & $0.57(14)$ & $2.401(64)$ & $0.2883(8)$\\
$5F_{7/2}$ & $0.71(20)$ & $2.610(75)$ & $0.415(12)$\\
$6S_{1/2}$ & $0.2029(59)$ & $0.1096(19)$ & $0.0673(11)$\\
$6P_{1/2}$ & $0.833(86)$ & $0.4304(93)$ & $0.1714(28)$\\ 
$6P_{3/2}$ & $0.99(32)$ & $0.4688(85)$ & $0.1987(31)$\\
\hline
\end{longtable}

$^a$\citep{pinnington1982beam}
$^b$\citep{curtis1989comprehensive}
$^c$\citep{bahr1982beam}

\twocolumn
\section{Results and Discussion}
\label{4}

The detailed analysis of our results is provided in this section along with comparison of our results with the available theoretical and 
experimental values from the literature. We have presented the calculated energies, line strengths, transition probabilities as well as oscillator 
strengths of the considered ions explicitly. We have also provided lifetimes of all the considered excited states from $4S$ through $6P_{3/2}$ for 
the above stated ions. The calculations for the E1 matrix elements are exclusively carried out by the relativistic AO method as discussed in 
previous section. It has been observed that our calculated E1 matrix elements using both the L- and V-gauge expressions are in good agreement 
for maximum number of transitions in all the considered ions. This, thereby, advocates for the reliability in our calculated results. It is, 
however, perceived that the energy values obtained using the SD method come out to be in better agreement with the experimental values than 
those from the SDpT method. It could be due to the fact that there may be large cancellations among the contributions arising from the non-linear 
terms of the SD approximation in the RCC theory and triple excitations as indicated in Refs. \citep{safronova2008all}. Thus, we consider E1 matrix
elements from the SD method to be more accurate and use them for estimating other properties. The differences between the results from 
the SD and SDpT methods are treated as possible uncertainties. The differences in the results obtained using the L- and V-gauge expressions are 
also included as uncertainties. In addition, a maximum of $1\%$ uncertainty from the calculated energies are taken into account in the results. 

Our calculated energies for a few low-lying states and their comparison with the available experimental values from the NIST AD 
\citep{ralchenko2008nist} are given in Table \ref{tab1}. As can be seen the deviation in energy values of all the considered states is less 
than $1\%$. We have observed maximum energy deviations of $0.15\%, 0.12\%$ and $0.53\%$ for the $4F_{7/2}, 5S_{1/2}$ and $4D_{5/2}$ states of 
As V, Se VI and Br VII ions respectively. This small differences between our calculated energies and experimental values indicate that the wave 
functions determined using the SD method are accurate and can be applied to estimate the radiative properties reliably. Hence, we expect a 
maximum of $1\%$ uncertainty in energies of high-lying states, which is therefore, included in our uncertainty calculations.

We have listed our results for wavelengths $\lambda$, line strengths $S_{vk}$, transition probabilities $A_{vk}$ and absorption oscillator 
strengths $f_{kv}$ for As V in Table \ref{tab2}. All these quantities are {\it ab initio}, and they are estimated by combining the calculated
energies and E1 matrix elements from the SD method. We have also presented the line strengths (square of the E1 matrix elements from both L- 
and V-gauge expressions using the SD wave functions, and observe a reasonable agreement between them. The uncertainties (quoted in the 
parentheses) for the $S_{vk}$, $A_{vk}$  and $f_{kv}$ values are estimated from the uncertainties of the E1 matrix elements as well as of energies.
Through the investigation of data of Table \ref{tab2}, it is accentuated that the spectroscopic properties of As V follows the systematic trend of
high transitions probabilities for transitions occurring between the low-lying states with the $4P_{3/2}$--$4D_{5/2}$ at the maxima. Small 
uncertainties are seen for the investigated properties in most of the allowed transitions except in the $4P_{1/2,3/2}$ -- $5D_{3/2,5/2}$, 
$4F_{5/2,7/2}$--$5D_{3/2,5/2}$, $5P_{1/2,3/2}$--$5D_{3/2,5/2}$, $5D_{5/2}$--$5F_{5/2,7/2}$ and $5D_{3/2,5/2}$--$6P_{1/2,3/2}$ transitions. This
may be because of the electron correlation contributions (difference between the DF and SD values) observed in the calculations of E1 matrix 
elements of these transitions. Nonetheless, the estimated uncertainties in these transitions are still reasonable for the applications of 
astrophysical studies.

The results for the wavelengths $\lambda$, line strengths $S_{vk}$, transition probabilities $A_{vk}$ and absorption oscillator strengths $f_{kv}$ 
for Se VI are presented in Table \ref{tab3}. It is seen that the $4F_{5/2}$--$5D_{5/2}$ transition exhibits lowest transition probability of 
order $3$, whereas the $4P_{3/2}$--$4D_{5/2}$ transition shows maximum transition probability of the order $10$. Therefore, we consider these
transitions to be least and most dominant transitions respectively, among all the $48$ considered transitions in Se VI. It is also analyzed that
there is a remarkably low uncertainty throughout the data except for the $4F_{5/2}$--$5D_{3/2,5/2}$ and $4F_{7/2}$--$5D_{5/2}$ transitions. The 
unreasonable large uncertainties of $23\%$ and $56\%$ are noticed in absorption oscillator strengths for $4F_{5/2}$--$5D_{3/2}$ and 
$4F_{7/2}$--$5D_{5/2}$ transitions respectively. This may be owing to the fact that there are about $11\%$ and $27\%$ difference between the 
L- and V-gauge matrix elements with respect to L-gauge results, respectively. Such large differences could be a consequence of strong 
correlations among the electrons from the $D$ and $F$ states. 

We have tabulated our spectroscopic data for Br VII in Table \ref{tab4}. Once again it is observed that the data follows the same trend as 
observed in case of the As V and Se VI ions, and $4P_{3/2}$--$4D_{5/2}$ is the most dominant transition. Relatively low uncertainty is noticed 
for this highly stripped ion in comparison to that of preceding Cu-isoelectronic ions. The maxima of uncertainty in the absorption transition
probability is located around 6\% for the $4D_{5/2}$--$5F_{5/2}$ transition.

Table \ref{tab6} demonstrates comparison of absorption oscillator strengths of a few transitions among the three ions with the available 
theoretical and experimental data. In this table, it is seen that the present results for transition probabilities and absorption oscillator 
strengths agree well with the reference data within quoted error limits for As V, Se VI and Br VII except for the results given by Engo et~al.
\citep{engo1997comparison}. This is expected because of entirely different approaches as well as the parameters involved in their estimations.
Our results deviate less than $17\%$ in comparison to the above stated reference \citep{engo1997comparison} except for the $4S$--$5P_{1/2,3/2}$, 
$4P_{1/2}$--$4D_{3/2}$ and $4P_{3/2}$--$4D_{3/2,5/2}$ transitions in As V. However, significant discrepancies have been noticed among the 
$4S$--$5P_{3/2}$ and $4P_{1/2}$--$4D_{3/2}$ transitions for Se VI; the corresponding deviations are about $48\%$ and $52\%$. Such disagreement 
between the $f_{kv}$ values is not seen for the Br VII ion when we compare of our results with respect to the theoretical data obtained by 
Migdalek and Baylis \citep{migdalek1979influence}, Curtis and Theodosiou \citep{curtis1989comprehensive} and Lavin et~al. 
\citep{lavin1994relativistic}. Migdalek and Baylis had used a relativistic model-potential with core-polarization (RMP-CP) effects for these 
calculations, while multi-configuration Dirac-Fock (MCDF) and RQDO methods were employed in Refs. \citep{curtis1989comprehensive} and 
\citep{lavin1994relativistic}, respectively. However, large discrepancies are seen with respect to the experimental values obtained using 
the beam-foil measurement technique by Knystautas and Drouin \citep{knystautas1977oscillator} {which are expected due to the fact that the beam-foil technique has the tendency to overestimate lifetimes and thus, underestimate oscillator strengths due to cascade effects attributed by high-lying levels}. This strongly suggests to carry out another 
independent measurement to affirm correctness of the above data. 

We have tabulated radiative lifetime values of the considered states and their comparison with the available literature data in Table \ref{tab5}.
It is realized that the radiative lifetime increases as we move on from the $S$ state to further excited states in general. However, for As V, the
$6P_{3/2}$ state exhibits maximum lifetime among all the considered states, whereas in case of Se VI, the $5D_{3/2,5/2}$ states show large 
lifetimes about $25.90$ ns and $55.96$ ns, respectively. We did not find such large lifetime for any state for As V and 
Br VII. In case of Br VII, the $5F_{7/2}$ state displayed large lifetime of $0.4151$ ns with an uncertainty less than $1\%$. During the comparison 
of lifetimes of a few low-lying states with available literature, it is realized that our results are in reasonable agreement with the results 
computed by Curtis and Theodosiou \citep{curtis1989comprehensive} using their MCDF method. Also, it is seen that the lifetime values remain within the quoted error 
limit with respect to the values given by Pinnington et~al. \citep{pinnington1982beam} as well as Bahr et~al. \citep{bahr1982beam}. Due to 
unavailability of data for the states above $5S$ in As V and Se VI, and above $4D_{5/2}$ in Br VII, we could not ascertain our results thoroughly 
and it calls for further theoretical and experimental studies. In case of Br VII, we believe that our aforementioned estimated values for various 
radiative properties are more reliable due to the fact that our employed AO method accounts electron correlation effects more rigorously than 
the previous estimations. Since earlier only a handful number of data were available and their uncertainties were not quoted, our reported values 
will be useful for the analyses of various astrophysical processes involving the Cu-isoelectronic As, Se and Br ions. Moreover, our precisely 
calculated values may be useful to guide the prospective experiments for their observations.

\section{Conclusion}
\label{5}

We have reported a large number of radiative properties of the Cu-isoelectronic As, Se and Br ions by employing an all-order many-body method 
in the relativistic coupled-cluster theory framework. This consists of precise estimations of line strengths, transition probabilities and 
absorption oscillator strengths of various transitions as well as the radiative lifetimes of many low-lying states. A total of $48$ transitions 
were considered for each of the three ions and the lifetimes are calculated for all the associated excited states from $4S$ to $6P_{3/2}$. We have also 
compared our results with the previously reported values for a few selected transitions and find a reasonably good agreement among them except 
in a couple of transitions in all the three ions. It requires further theoretical and experimental investigations to affirm correctness of these 
data. Our data with the quoted uncertainties can provide a benchmark for many applications in the astrophysical processes and for their 
observations in the future.

\section*{Acknowledgement}

The work of B. A. is supported by SERB-TARE research grant no. TAR/2020/000189, New Delhi, India. The employed all order method was developed in the group of Professor M. S. Safronova of the University of Delaware, USA.

\section*{Data availability}

The data underlying this article are available in the article.

\bibliographystyle{mnras}

\end{document}